\documentclass[preprint,showpacs,preprintnumbers,amsmath,amssymb]{revtex4}
\usepackage{amsmath,amssymb,mathrsfs,amsthm}
\usepackage{dsfont}
\usepackage{color}
\usepackage{graphics}
\usepackage{graphicx}
\usepackage{dcolumn}
\usepackage{bm}
\usepackage{epsfig}

\newcommand{\Res}{{\rm Res}}

\begin{document}
\sloppy
\title{Universal low-energy physics in 1D strongly repulsive multi-component Fermi gases } 

\author{Yuzhu Jiang}

\affiliation{${1}$ State Key Laboratory of Magnetic Resonance and
Atomic and Molecular Physics, Wuhan Institute of Physics and
Mathematics, Chinese Academy of Sciences, Wuhan 430071, China}

\author{Peng He}
\affiliation{${2}$ Bureau of Frontier Sciences and Education, Chinese Academy of Sciences, Beijing 100190, P. R. China}

\author{Xi-Wen Guan}
\email[e-mail:]{xiwen.guan@anu.edu.au}
\affiliation{State Key Laboratory of Magnetic Resonance and Atomic and Molecular Physics,
Wuhan Institute of Physics and Mathematics, Chinese Academy of Sciences, Wuhan 430071, China}
\affiliation{Center for Cold Atom Physics, Chinese Academy of Sciences, Wuhan 430071, China}
\affiliation{Department of Theoretical Physics, Research School of Physics and Engineering,
Australian National University, Canberra ACT 0200, Australia}

\begin{abstract}

It was shown [Chin. Phys. Lett. 28, 020503 (2011)] that at zero temperature the ground state of the  one-dimensional (1D) $w$-component Fermi gas coincides with that of the spinless Bose gas in the limit $\omega\to \infty$. 
This behaviour  was experimentally  evidenced through a quasi-1D  tightly trapping ultracold  ${}^{173}$Yb atoms in the recent paper [Nature Physics 10, 198 (2014)].
However, understanding of low temperature behaviour of the Fermi  gases with a repulsive interaction acquires spin-charge separated conformal field theories of an
effective Tomonaga-Luttinger liquid and an antiferromagnetic $SU(w)$ Heisenberg spin chain.
Here we analytically derive  universal thermodynamics of 1D strongly repulsive fermionic gases with $SU(w)$ symmetry via the Yang-Yang thermodynamic Bethe ansatz method.
The analytical free energy and magnetic properties of the systems at
low temperatures in a weak magnetic field are obtained  through  the
Wiener-Hopf method.
In particular, the free energy essentially manifests   the spin-charge separated conformal field theories  for the high symmetry  systems with arbitrary repulsive interaction strength.
We also find  that the sound velocity of the  Fermi gases in the large $w$ limit coincides with that for the spinless Bose gas, whereas the spin velocity vanishes quickly as $w$ becomes large.
This indicates a strong  suppression of the Fermi exclusion statistics by the commutativity feature among the $w$-component fermions with different spin states in the Tomonaga-Luttinger liquid phase.
Moreover, the equations of state and critical behaviour of physical quantities at finite temperatures are analytically derived in terms of the polylogarithm functions in the quantum critical region.
%
%
\end{abstract}


\maketitle

\tableofcontents

\section{Introduction}

Advances in manipulating alkaline-earth atoms provide  a promising platform for studying quantum systems  with high spin symmetries \cite{Gorshkov2010NP, Gorshkov2014NP,Bloch2008RMP, XWGuan2013RMP, TLHo1998PRL, Wu:2003,Bloch2012NP,Krauser2012,Cazalilla2014}.
The nuclear spin of most alkaline-earth isotopes is either zero or half-odd number, therefore the ultracold gases of these atoms are either single component bosons or multicomponent fermions.
The electronic spins decouple from nuclear spins in alkaline-earth atoms that gives $SU(w)$ symmetry with $w=2I+1$ in two-body collisions, here $I$ is the nuclear spin.
Through fully controllable interaction and spin states, a variety of high symmetry systems have been realized in the laboratories.
 For example, ${}^{173}$Yb and ${}^{87}$Sr are the alkali earth fermionic atoms  which can display high spin symmetries, i.e. 
 $^{173}$Yb atoms have $SU(6)$ symmetry \cite{Fukuhara2007PRL, Taie2010PRL} and $^{87}$Sr  atoms have $SU(10)$ symmetry \cite{XZhang2014S}.
These experimental developments provide exciting opportunities to explore a wide range many-body phenomena such as spin and orbital magnetism \cite{Cappellini2014PRL, Scazza2014NP}, Kondo spin-exchange physics \cite{Nakagawa:2015,Zhang2015}, the synthetic dimension \cite{Zeng2015} and  the one-dimensional (1D) Tomonaga-Luttinger liquid (TLL) \cite{Pagano2014NP} etc.
Ultracold alkaline-atoms have opened many ways to study low-dimensional systems of interacting fermions, bosons and spins, see recent review papers \cite{YALiao2010N,CazalillaRMP,XWGuan2013RMP,Murray2015}.
Large spin systems of cold atoms exhibit rich internal structures  which may result  in multi-component quantum liquids and diverse   critical phenomena \cite{XWGuan2013RMP,Cazalilla2014,YCYu2015,Wu:2006,Capponi:2015,Schlottmann:1997}.
Theoretical progress toward understanding large spin magnetism and quantum liquids  was made particular on  $SO(5)$ symmetry and $SU(w)$ symmetry in the scenario of alkaline-earth atoms.
From the ultracold atom perspective, the integrable model of the $SU(w)$ Fermi gases, which  was solved  by Sutherland \cite{Sutherland1968PRL}  in 1968, are having high impact, see review  \cite{XWGuan2013RMP}.
In this regard,  some new integrable  systems with exotic symmetries such as $SO(w)$ and $Sp(w)$ symmetries have been recently constructed  by using  the Bethe ansatz (BA)  \cite{JPCao2007EPL,Jiang2011JPA}.
So far the exact results of 1D integrable quantum systems with large symmetries have provided  a better  understanding of  large  spin magnetism, quantum liquids,   universal  thermodynamics and universal laws \cite{XWGuan2013RMP,Cazalilla2014,YCYu2015,Schlottmann:1997,Guan2010PRA,GuanXW2011PRA,GuanIJMP,Patu:2015}.

In regard of the integrability, the renewed interest over the past decade has been paid in  the exactly solved models of interacting fermions and bosons.
Since the pioneering work in the 60's, 70's and 80's of McGuire,
Yang, Lieb, Sutherland, Baxter \emph{et al.},  for example  \cite{McGuire:1964,Lieb1963PR, LiebLiniger1963PR, Yang1967PRL,Baxter:1972},  the  understanding of integrable
models has greatly extended our knowledge in the theories of many-body physics, condensed matter physics, ultracold atomic physics,
quantum phase  transitions and critical phenomena.
The 1D many-body systems exhibit rich   properties  some of which might be  significantly different  from that of the models in  higher dimensions.
In particular, the  hallmark of 1D many-body physics  is the  spin-charge separation phenomenon \cite{Giamarchi:2004,1D-Hubbard,CazalillaRMP,Recati2003PRL, Fuchs2005PRL, Kollath2005PRL} which does not exist in higher dimensions.
In this scenario, 1D integrable quantum systems, solved by means of the  BA method \cite{Lieb1963PR, LiebLiniger1963PR, Yang1967PRL, Gaudin1967PLA}, usually provide  a rigorous proof  of  such unique 1D many-body  phenomenon.
Affleck
\cite{Affleck1986} and Cardy \cite{Cardy1986}  showed that conformal invariance gives a
universal forms for the finite temperature and finite size effects in 1D systems.
In this context, Mezincescu \emph{et al.}
\cite{Nepomechie1992,Nepomechie1993} derived the finite temperature correction  in  the free energy of
spin chains  under a small magnetic field by
using the Wiener-Hopf technique, also see \cite{Schlottmann:1997}.
Following these methods, analytical free energies  and magnetic properties for the two- and three-component repulsive Fermi gases  \cite{Lee2012PRB,PHe2011JPA}  at low temperatures have been derived from the thermodynamic Bethe ansatz (TBA) equations,    \cite{Yang1969JMP,Takahashi1999,XWGuan2013RMP,Schlottmann:1997}.
Recently,  the quantum transfer matrix method has been adapted to treat thermodynamics of the 1D  interacting Bose and Fermi gases \cite{Patu:2015,Klumper:2011,Patu:2015b}.
Despite much theoretical study  on the  1D  Fermi gases in literature, rigorous  derivation of   such unique 1D many-body phenomenon for these high symmetry systems  still has   not been achieved.
In this paper, we  aim  to investigate universal feature of the TLL in the 1D $SU(w)$ symmetry Fermi gases with repulsive interactions by solving  the TBA equations.
The analytical free energy and magnetic properties of the systems at low temperatures in a weak magnetic field are derived by using  the Wiener-Hopf method.
At low temperatures,  the free energy of the systems shows  the spin-charge separated conformal field theories  of an
effective TLL and an antiferromagnetic $SU(w)$ Heisenberg spin chain, where the central charge in the charge sector $C_{\rm c}=1$ and the central charge in the spin sector is $C_{\rm s}=w-1$ in this weak magnetic field limit.
A general relation between the magnetic field and the spin velocity under pure Zeeman splitting is obtained. 
We find that the sound velocity  of the  Fermi gases in the large $w$ limit coincides with that of  the spinless Bose gas, whereas the spin velocity vanishes  quickly as $w$ becomes  large.
This nature gives rise to a precise  understanding of the experimental observation \cite{Pagano2014NP} that at low temperature the $w$-component Fermi gases display  the bosonic spinless  liquid for a large value of $w$.
Our finding is consistent with the ground state properties of the high symmetry Fermi gases which were observed in \cite{Yang:2011,GMW:2012}.
Furthermore, we study  the thermodynamics and quantum criticality of the systems  beyond the regime of the Tomonaga-Luttinger liquid phase.
Our result reveals the universal behaviour  of  interacting fermions with high symmetries in 1D.

This paper is organized as follows.
In Section \ref{sec-MTBA},  the BA equations for 1D $SU(w)$ quantum gases are  presented.
In Section \ref{Sec-LTWM}, the Sommerfeld expansion of TBA equations in  weak magnetic field and low temperature regime is carried out.
In section \ref{sec-WH},
using  the Wiener-Hopf method, we obtain the low temperature free energy.
In section \ref{sec-LTSCS}  we discuss  the spin-charge separation phenomenon.
In Section \ref{sec-SR}, we discuss the TLL and critical behaviour of the strong coupling Fermi gases.
Section \ref{sec-C} is reserved as our  conclusion and discussion.

\section{The Model and thermodynamic Bethe ansatz equations}
\label{sec-MTBA}
We consider a 1D  $\delta$-function interacting  $w$-component fermionic  system of $N$ particles with mass $m$, where the interactions between different components have the same coupling strength $g_{\rm 1D}$.
%
%
The Hamiltonian of this system reads \cite{Sutherland1968PRL}
\begin{align}
 \label{H}
 H=-\frac{\hbar ^{2}}{2m}\sum_{i=1}^{N}\frac{\partial^2}{\partial x_i^2} +g_{1D}\sum_{i<j}\delta(x_{i}-x_{j})+h \hat S_z,
\end{align}
where $\hat S_z$ is the total spin of the $z$-direction, $\hat S_z=
\sum_{r=1}^{w}[-(w+1)/2+r )]N_{r}$ and $h$ is the external magnetic field. Here $N_r$ is the particle number in the hyperfine  state $r$.
There are $w$ possible hyperfine states $|1\rangle, |2\rangle, \ldots,
|w\rangle$ that the fermions can occupy.
For the spin independent interactions, the number of fermions  in each spin state is  conserved.
In the above equation, the interaction coupling constant is given by $g_\mathrm{1D} =-2\hbar^2/m a_\mathrm{1D}$, here  $a_\mathrm{1D}$ is the effective scattering length in 1D \cite{Ols98}.
The interaction between fermions in different spin states is repulsive when $g_{\rm 1D}>0$ and attractive when $g_{\rm 1D}<0$.
Following the BA convention, we also introduce the interaction strength $c=m g_{1D}/\hbar^2$.
 From now on, we shall choose our units such that $\hbar^2=2m=1$ unless we particularly use the units.

The Hamiltonian (\ref{H}) has the symmetry of $U(1)\times SU(w)$ when the magnetic field is absent, where $U(1)$ and $SU(w)$ are the symmetries of the charge and spin degrees of freedom, respectively.
Although the Zeeman splitting   breaks  the $SU(w)$ symmetry, $\hat S_z$ is a conserved quantity.  The model can be solved exactly  via Bethe ansatz (BA) \cite{Sutherland1968PRL} using the approach proposed by \cite{Yang1967PRL,Yang1968PR}.
In the following calculation, we assume that the system is constrained to a line with a length $L$.
The energy is given by
\begin{align}
 E=\sum_{j=1}^Nk_j^2,
\end{align}
where the pseudo-momenta $\left\{k_j\right\}$ are determined by the BA equations,
\begin{align}
 \label{BAE}
 & {\rm e}^{\mathrm{i}k_{j}L}=\prod_{\ell=1}^{M_{1}}\frac
 {k_{j}-\lambda^{(w-1)}_{\ell}+{\rm i}\frac c2}
 {k_{j}-\lambda^{(w-1)}_{\ell}-{\rm i}\frac c2},
 j=1,\ldots,N.
 \nonumber \\
 &
 \prod_{j=1}^{N}\frac
 {\lambda^{(w-1)}_{\ell}-k_{j}+{\rm i}\frac c2}
 {\lambda^{(w-1)}_{\ell}-k_{j}-{\rm i}\frac c2}
 \prod_{m=1}^{M_{w-2}}\frac
 {\lambda^{(w-1)}_{\ell}-\lambda^{(w-2)}_{m}+{\rm i}\frac c2}
 {\lambda^{(w-1)}_{\ell}-\lambda^{(w-2)}_{m}-{\rm i}\frac c2}\nonumber\\
 &=-\prod_{\alpha =1}^{M_{w-1}}\frac
 {\lambda^{(w-1)}_{\ell}-\lambda^{(w-1)}_{\alpha}+\mathrm{i}c}
 {\lambda^{(w-1)}_{\ell}-\lambda^{(w-1)}_{\alpha}-\mathrm{i}c},
 \ell =1,\ldots ,M_{w-1},
 \nonumber \\
 &
 \prod_{j=1}^{M_{r+1}} \frac
 {\lambda^{(r)}_{\ell}-\lambda^{(r+1)}_{j}+{\rm i}\frac c2}
 {\lambda^{(r)}_{\ell}-\lambda^{(r+1)}_{j}-{\rm i}\frac c2}
 \prod_{m=1}^{M_{r-1}}\frac
 {\lambda^{(r)}_{\ell}-\lambda^{(r-1)}_{m}+{\rm i}\frac c2}
 {\lambda^{(r)}_{\ell}-\lambda^{(r-1)}_{m}-{\rm i}\frac c2}\nonumber\\
 &=-\prod_{\alpha =1}^{M_{r}}\frac
 {\lambda^{(r)}_{\ell}-\lambda^{(r)}_{\alpha}+\mathrm{i}c}
 {\lambda^{(r)}_{\ell}-\lambda^{(r)}_{\alpha}-\mathrm{i}c},~~
 r=2,3,\cdots,w-2,
 \ell =1,\ldots ,M_{r},
 \nonumber \\
 &
 \prod_{j=1}^{M_2} \frac
 {\lambda^{(1)}_{\ell}-\lambda^{(2)}_{j}+{\rm i}\frac c2}
 {\lambda^{(1)}_{\ell}-\lambda^{(2)}_{j}-{\rm i}\frac c2}
 =-\prod_{\alpha =1}^{M_1}\frac
 {\lambda^{(1)}_{\ell}-\lambda^{(1)}_{\alpha}+\mathrm{i}c}
 {\lambda^{(1)}_{\ell}-\lambda^{(1)}_{\alpha}-\mathrm{i}c},
 \ell =1,\ldots ,M_1.
\end{align}
Here $\{\lambda^{(r)}\}$, $r=1,2,\cdots,w-1$ denote the spin rapidities which are introduced  to describe the motion of spin waves. The particle number $N_r$ in each spin state links to the quantum number $M_\alpha$  via the relation $N_r=M_{r}-M_{r-1}$ and $M_0=0$.
In this paper, we will consider the repulsive interaction, i.e.  $c>0$.
There is no charged bound state for  the repulsive interaction, and
the pseudo-momenta $\left\{k_{j}\right\}$ are hence real.
At the thermodynamic limit, i.e. $N, L\rightarrow \infty$ and $n=N/L$ is finite.
Each branch of spin rapidities $\{\lambda^{(r)}\}$ has complex roots
\begin{align}
 \lambda^{(r)}_{q,j,z} &= \lambda^{(r)}_{q,j}
 -\frac{\mathrm{i} c }{2}(q+1-2 z),\qquad z = 1,\cdots,q,
\end{align}
at the thermodynamic limit.
For given $r,q$ and $j$, rapidities with different $z$ share the same real part $\lambda^{(r)}_{q,j}$, which are called $q$-string in the spin branch $r$.
Here the number $q$ is the length of the string, and $j=1,2,\cdots,M^{(r)}_q$ label the different real parts of the $q$-strings, where $M^{(r)}_q$ is the number of the $q$-strings with $M^{(r)}=\sum_q q M^{(r)}_q$.

In order to carry out our calculation, we first recall some basis notations for the TBA equations. At the thermodynamic limit, the particle densities $\{\rho\}$, hole densities $\{\rho_{\rm h}\}$ are introduced to describe the equation of the  state of system and the dressed energies $\varepsilon$ are defined as $\varepsilon=T\ln(\rho_{\rm h}/\rho)$.
Following Yang-Yang's grand canonical method \cite{Yang1969JMP}
 the TBA equations for  the model (\ref{H}) are given by \cite{Takahashi1999,XWGuan2013RMP,Schlottmann:1997,Lee:2011}
\begin{eqnarray}
 \label{TBAE}
  \textstyle
 \varepsilon^{\rm c}(k)&= &k^{2}-\mu -sh
 +\sum_{n=1}^{\infty}\hat a_{n}\ast \varepsilon^{w-1,n}_-,
 \nonumber\\
  \textstyle
 \varepsilon^{w-1,n}(\lambda)&=&nh
 +\hat a_{n}\ast \varepsilon^{\rm c}_-
 +\sum_m \hat C_{n,m}\ast \varepsilon^{w-2,m}_-
 -\sum_m \hat T_{mn}\ast \varepsilon^{w-1,m}_-,
 \nonumber\\
  \textstyle
 \varepsilon^{r,n}(\lambda)&=&nh
 +\sum_m \hat C_{n,m}\ast (\varepsilon^{r-1,m}_-+\varepsilon^{r+1,m}_-)
 -\sum_m \hat T_{mn}\ast \varepsilon^{r,m}_-,
 \nonumber\\
 \textstyle
 \varepsilon^{1,n}(\lambda)&=&nh
 +\sum_m \hat C_{n,m}\ast \varepsilon^{2,m}_-
 -\sum_m \hat T_{mn}\ast \varepsilon^{1,m}_-.
\end{eqnarray}
Where $\varepsilon^{\rm c}(k)$ and $\varepsilon^{r,n}(k)$ are the dressed energies for  the charge sector and for  the branch $r$ in  the spin sector, respectively and  $\varepsilon_\pm= T\ln(1+{\rm e}^{\pm \varepsilon/T})$.  As  a convention used in the TBA equations,  $n$ labels  the length of the strings and  we denote $\ast$ as the convolution integral, i.e. $\hat a\ast f(x)=\int_{-\infty}^{\infty}
a(x-y)f(y){\rm d}y$ and the integral kernels are given by
\begin{eqnarray}
 a_{m}(x)& =&\frac{1}{2\pi }\frac{mc}{(mc/2) ^{2}+x^{2}},
 \nonumber\\
 T_{mn}&=& a_{m+n}+2a_{m+n-2}+\ldots+2a_{|m-n|+2}+(1-\delta_{nm})a_{|m-n|},
 \nonumber\\
 C_{mn}&=&a_{m+n-1}+a_{m+n-3}+\ldots+a_{|m-n|+3}+a_{|m-n|+1}.
\end{eqnarray}

Using the Fourier transformation,  we may rewrite the above TBA equations (\ref{TBAE})   into the following recursive  form
\begin{eqnarray}
 \label{TBAr}
 \textstyle
 \varepsilon^{
 \rm c} &=& k^2-\mu
 + \hat G_{\rm c} \ast \varepsilon^{\rm c}_-
 - \sum_{r=1}^{w-1}\hat G_{r}\ast \varepsilon^{r,1}_+ , \nonumber  \\
 \textstyle
 \varepsilon^{r,n} &=& \hat G\ast
 (\varepsilon^{r+1,n}_++\varepsilon^{r-1,n}_+
 +\varepsilon^{r,n+1}_++\varepsilon^{r,n-1}_+),
\end{eqnarray}
where $\varepsilon^{r,0}_\pm=\varepsilon^{0,n}_\pm=\varepsilon^{w,m}_\pm=0$ when $m>1$ and
$\varepsilon^{w,1}=\varepsilon^{\rm c}$ and the integral kernels are
\begin{eqnarray}
 \label{kes2}
 G_r(x)&=&\frac{(wc)^{-1} \sin (r\pi/w)}{ \cos(r\pi/w)+\cosh(2\pi x/wc)},~~r=1,2,\cdots,w-1,\\
 G_{\rm c}&=&\hat G_{w-1}\ast a_1,~~~G(x)= \frac{1}{2 c \cosh \left(\pi x /c \right)},
\end{eqnarray}
The boundary condition of these recursive equations is
\begin{align}
 \lim_{n\rightarrow \infty} \varepsilon^{r,n} =nh.
 \label{limit}
\end{align}

\begin{figure}[ht]
 \includegraphics[width=0.9\linewidth]{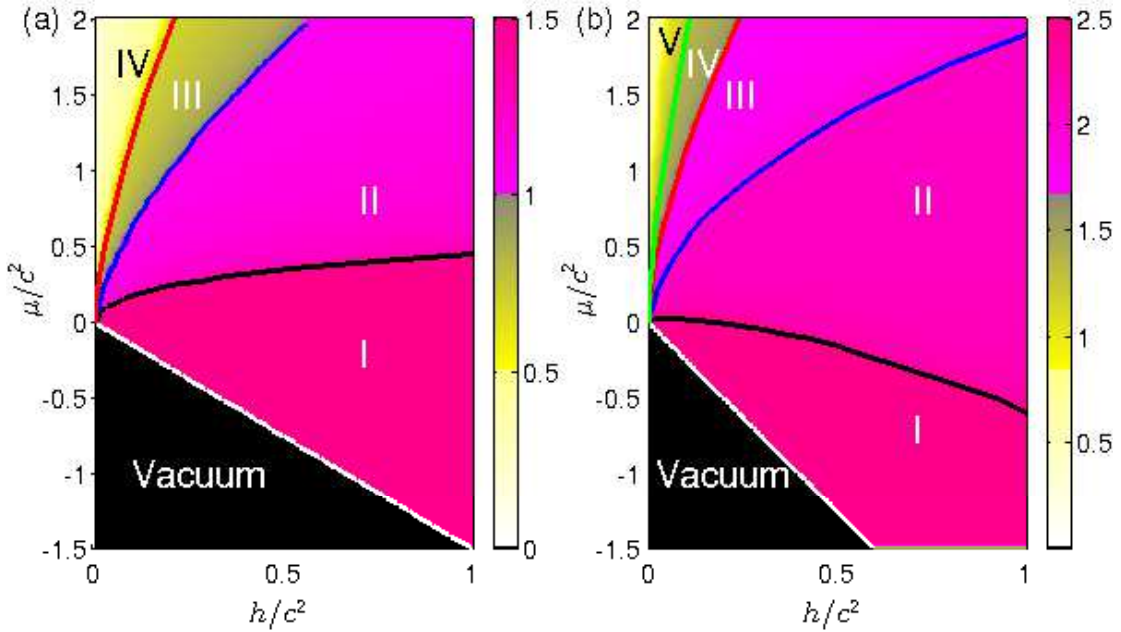}
 \caption{
 \label{fig-pd}
 Phase diagram of 1D repulsive $SU(w)$ fermions: (a) The phase diagram of spin-$3/2$ model. (b) The phase diagram of spin-$5/2$ model. See the main  text.}
\end{figure}

The TBA equations (\ref{TBAE}) determine  full thermodynamics and critical behaviour of the 1D $SU(w)$ quantum gases. For zero temperature, i.e. $T\to 0$, the model  exhibits   a very rich phase diagram. In this limit, the TBA equations become the linear integral equations from which the full phase diagram can be derived analytically and numerically.
When the external magnetic field $h$ is stronger than the saturation field $h_c=\mu/s$,  the system is fully  polarized to  the state with spin $s_z=s=(w-1)/2$.
For decreasing the magnetic field $h$, Zeeman splitting become weaker. Thus  different branches of spin states emerge in the ground  state.
Solving the TBA equations (\ref{TBAE}) in the zero temperature limit, we plot the phase diagrams of the  spin-$3/2$ and spin-$5/2$ Fermi gases in Fig. \ref{fig-pd}.  Where  phase (I) stands for  the fully polarized state where $s_z=s$.
phase (II) is the mixture of two spin states $s_z=s$ and $s_z=s-1$, phase (III) denotes  the mixture of three branches  where $s_z=s$, $s_z=s-1$ and $s_z=s-2$ and so on.

\section{Low temperature and  weak magnetic field regime }
\label{Sec-LTWM}

For  the ground state, i.e. $T=0$, all  spin rapidities are real, thus we can denote $\varepsilon^{r,1}=\varepsilon^{r}$ for our convenience.
At finite temperatures, the string solutions (complex solutions) emerge in each spin branches.
Analyzing the TBA equations  (\ref{TBAE}) or (\ref{TBAr}), very small  numbers of strings  involve in the low temperature thermodynamics.
The presence of these strings make the nonlinear TBA equations very hard to  be solved analytically.
However, when the temperature is much smaller than magnetic field, $T\ll h$, the TBA equations of dressed energies can be simplified as
\begin{eqnarray}
 \label{TBAg}
 \varepsilon^{\rm c} &= & k^2-\mu+ \hat G_{\rm c}\ast \varepsilon^{\rm c}_-
 - \sum_{r=1}^{w-1}\hat G_{r}\ast \varepsilon^{r}_+ , \nonumber  \\
 \varepsilon^{r} &=& h+\hat G_r\ast \varepsilon^{\rm c}_-+\sum_{r'}\hat R_{rr'}*\varepsilon^{r'}_+,~~r=1,2,\cdots,w-1
 \end{eqnarray}
 up to ignorance of higher length  strings, see the method used in  \cite{Lee2012PRB,PHe2011JPA}.
For  the ground state, $\varepsilon^{r}(\pm \infty)\geq0$ and $\varepsilon^{r}(0)<0$,
which means $\varepsilon^{r} (k)$ has two zero points $\pm Q_r$ in the spin rapidity space of the $r$-branch, i.e. two Fermi like points in the spin rapidity space.
For zero  external magnetic field, the zero point $Q_r\to \infty$. For  a weak magnetic field $h$,  $Q_r$ is  very  large. 
%
%
If the magnetic field $h$ is very small, i.e. $hc\ll p$, we can use the Wiener-Hopf method to solve the second equation of (\ref{TBAg}) in spin sectors. 
Here the pressure is defined by
\begin{align}
 \label{p}
 p=-\frac{1}{2\pi}\int_{-\infty}^{\infty} \varepsilon^{\rm c}_-(k) {\rm d}k.
\end{align}

Before further going on, we define some  convolution operators: $\circ$ is defined by
$\hat R\circ \varepsilon(k)=
 \big[\int_{-\infty}^{-Q}
 +\int_{Q}^{\infty}\big] {\rm d}k'R(k-k') \varepsilon(k')$ and $\bullet$ is defined by
$\hat R\bullet \varepsilon(k)=
 \int_{-Q}^{Q} {\rm d}k'R(k-k') \varepsilon(k')$. When $T\ll|\varepsilon(0)|$, we can apply Sommerfeld expansion technique to eq. (\ref{TBAg}) with the temperature $T\rightarrow 0$.
We consider the leading order
\begin{eqnarray}
 R\ast \varepsilon_+(k)\approx R\circ \varepsilon(k) + R(k,Q)\xi,\,\,\,R\ast \varepsilon_-(k)\approx R\bullet  \varepsilon -R(k,Q)\xi \nonumber
\end{eqnarray}
where $R(k,Q)=R(k+Q)+R(k-Q)$ and $\xi=\pi^2T^2/6\varepsilon'(Q)$.
The equations dressed energies are thus  given by
\begin{align}
 \label{tba-t}
 &
 \varepsilon^{\rm c} = k^2-\mu
 + \hat G_{\rm c}\bullet \varepsilon^{\rm c}
 - \sum_{r=1}^{w-1}\hat G_{r}\circ \varepsilon^{r}
 - G_{\rm c}(k,Q_{\rm c})\xi_{\rm c}
 - \sum_{r=1}^{w-1} G_{r}(k,Q_r)\xi_r,\\
 &
 \varepsilon^r = h
 +\hat G_r\bullet\varepsilon^{\rm c}
 +\sum_{r'}\hat R_{rr'}\circ \varepsilon^{r'}
 -G_r(k,Q_{\rm c}) \xi_{\rm c}
 +\sum_{r'} R_{rr'}(k,Q_{r'})\xi_{r'}.
 \nonumber
\end{align}
When the magnetic field is very small, the zero point $Q_r$ will be so  large such that the quantity  $z={\rm e}^{-Q_r \pi/c}\ll 1$.
Up to the first order of $z$, we get the following asymptotic forms:
\begin{align}
 &
 \hat G_r\circ \varepsilon^{r}(k)
 =\frac{4\sin(r\omega_0 c)}{wc}\cosh(\omega_0k) {\rm e}^{-{\rm i}\omega_0 Q_r}
 \int_0^\infty {\rm d}k' \varepsilon^r(k'+Q_r),
 \nonumber\\
 &
 \hat R_{rr'}\circ\varepsilon^{r'}(k+Q_{r'})
 =\int_0^\infty {\rm d}k' \hat{\mathds{R}}_{rr'}(k-k')
 \varepsilon^{r'}(k'+Q_{r'})+O(z^2),
 \nonumber\\
 &
 \hat G_r\bullet\varepsilon^{\rm c}(k)
 =b_\varepsilon \hat G_r(k),~~~~r,r'=1,2,\cdots,w-1,~~~\omega_0=2\pi/wc,
 \nonumber\\
 &
 \label{b}
 b_\varepsilon
 =\hat I\bullet \big[\cosh(\omega_0 k) \varepsilon^{\rm c}(k) \big],
\end{align}
where $\hat I$ is the convention operator of kernel $\delta(x-x')$.
Submit these results into eq. (\ref{tba-t}), we have
\begin{eqnarray}
 \label{tba-thp1}
 \varepsilon^{\rm c} &=& k^2-\mu
 + \hat G_{\rm c}\bullet\varepsilon^{\rm c}(k)
 - G_{\rm c}(k,Q_{\rm c})\xi_{\rm c}
 - \sum_{r=1}^{w-1} G_{r}(k,Q_r)\xi_r \nonumber\\
 &&
 - \sum_{r=1}^{w-1}
 \frac{4\sin(r\omega_0 c)}{wc}\cosh(\omega_0k) {\rm e}^{-{\rm i}\omega_0 Q_r}
 \int_0^\infty {\rm d}k' \varepsilon^r(k'+Q_r),
 \\
 \varepsilon^r(k+Q_r) &=& h
 +b_\varepsilon G_r(k+Q_r)
 +\sum_{r'}\int_0^\infty \hat{\mathds{R}}_{rr'}\circ \varepsilon^{r'}(k+Q_{r'})
 \nonumber\\
 &&
 -2\cosh(\omega_0Q_{\rm c})G_r(k+Q_r) \xi_{\rm c}
 +\sum_{r'} \mathds R_{rr'}(k+Q_r,Q_{r'})\xi_{r'}.
 \nonumber
\end{eqnarray}

In  the second equation  of (\ref{tba-thp1}),  we may take the following approximations  $R_{rr'}(k+Q_r,Q_{r'})\approx \mathds{R}_{rr'}(k)$ and  $G_{r}(k+Q_r,Q_{\rm c})\approx 2\cosh(\omega_0 c)G_{r}(k+Q_r)$.
We define functions $y_\pm$ as 
\begin{align}
 y^r_{\pm}(k)=\theta(\pm k)\varepsilon^{r}(k+Q_{r}),
\end{align}
where $\theta(k)$ is the step function, $\theta(k)=1$ when $k>0$ and $\theta(k)=0$ when $k<0$.
Submitting the function $ y^r_{\pm}$   into the equation of $\varepsilon^{r}$, we obtain
the Wiener-Hopf type equation
\begin{align}
 &\label{wh-wh}
 y^r_+(k)+y^r_-(k)=g_r(k)+\sum_{r'} \mathds{R}_{rr'}*y^r_{+}(k),
\end{align}
where $y^r_+(k)$  relates to  the dressed energy $\varepsilon^{r}(k+Q_{r})$ for  $k>0$.  Whereas $y^r_-(k)$ is just a continuation of $y^r_+(k)$ via  eq.(\ref{wh-wh}).
Up to the first order of $z$, the $g_r$ function is
\begin{eqnarray}
 \label{g}
 g_r(k)&=&h+B_\varepsilon\hat G_r(k+Q_r)+\sum_{r'} \mathds{R}_{rr'}(k)\xi_{r'},\\
\label{B}
 B_\varepsilon
 &=&b_\varepsilon-2\cosh(\omega_0 Q_{\rm c})\xi_{\rm c}.
\end{eqnarray}
We will solve the eq. (\ref{wh-wh}) to get $y^r_+(k)$ with the help of the  Wiener--Hopf method  next section.

For charge sector, the equation  (\ref{tba-t}) can be rewritten as
\begin{align}
 \label{tba-thp}
  &
 \varepsilon^{\rm c} = k^2-\mu
 + \hat G_{\rm c}\bullet\varepsilon^{\rm c}(k)
 - G_{\rm c}(k,Q_{\rm c})\xi_{\rm c}
 - \sum_{r=1}^{w-1} G_{r}(k,Q_r)\xi_r \nonumber\\
 &\quad~
 - \sum_{r=1}^{w-1}
 \frac{4\sin(r\omega_0 c/2)}{wc}\cosh(\omega_0k) {\rm e}^{-\omega_0 Q_r}
 \int_0^\infty {\rm d}k' \varepsilon^r(k'+Q_r).
\end{align}
Substituting  the function $y^r_+(k)$ into the eq.(\ref{tba-thp}), we further obtain
\begin{align}
 \label{tba-whp}
 &
 \varepsilon^{\rm c} \approx k^2-\mu
 + \hat G_{\rm c}\bullet\varepsilon^{\rm c}(k)
 - G_{\rm c}(k,Q_{\rm c})\xi_{\rm c}
 -4\cosh(\omega_0 k)S_\varepsilon/wc,
 \nonumber\\
 & 
 S_\varepsilon=\sum_r \sin(\omega_0 rc/2) {\rm e}^{-\omega_0 Q_r}
 \big[\tilde y_{+}({\rm i}\omega_0)+\xi_r\big],
\end{align}
where $\tilde y(\omega)$ is defined by the Fourier transform
\begin{align}\label{FT}
 \hspace{-10pt}
 \tilde f(\omega) = \int_{-\infty}^\infty
 f(k) {\rm e}^{{\rm i} k \omega} {\rm d}k.
\end{align}

\section{Wiener--Hopf solution}
\label{sec-WH}

By using the Wiener--Hopf method we will solve  the eq.(\ref{wh-wh}) in this section.
Applying  the Fourier transform (\ref{FT}) to  eq.(\ref{wh-wh}),  we have
\begin{align}
 \label{wh-FT0}
 \tilde y^r_{+}+ \tilde y^r_{-}=\tilde g_r + \sum_{r'} \tilde{R}_{rr'} *\tilde y^r_{+}.
\end{align}
If the functions $y^r_\pm(\pm k)$ in eq.(\ref{wh-wh}) are analytic and bounded for  $k>0$, $\tilde y_\pm(\omega)$ are analytic in the upper/lower half plane with the condition $\tilde y_\pm(\infty)=0$.

In the framework of Wiener-Hopf method, the integral kernels $(1-R)^{-1}$ must be  factorized into two functions which are analytic in upper/lower half plane.
However, the Eq.(\ref{wh-FT0}) is difficult to solve for the case that  $R$ is a matrix.
The $R$ can be diagonalized by using the transformation  matrix
\begin{align}
 U_{rr'}=\sqrt{\frac{2}{w}}\sin\frac{rr'\pi}{w},
\end{align}
where $U$ is a unitary matrix, $UU=1, U=U^{-1}, U=U^t, U=U^\dag$.
Making the following transformation,
\begin{align}
 &
 \textstyle
 \bar y^r_\pm =\sum_{r'} U_{rr'} \tilde y^r_\pm,~
 \bar g_r =\sum_{r'} U_{rr'} \tilde g_r,~
 \bar R_r\delta_{rr'} =\sum_{r_1r_2} U_{rr_1} \tilde R_{r_1r_2} U_{r_2r'},\\
 &
 (1-\bar R_r)^{-1} =4 \tilde a_1
 \cosh\Big[\frac{\omega c}4-\frac{\pi{\rm i}}{2}\frac{r-w}{w}\Big]
 \cosh\Big[\frac{\omega c}4+\frac{\pi{\rm i}}{2}\frac{r-w}{w}\Big],
 \nonumber
\end{align}
we then obtain  the diagonalized Wiener-Hopf equation
\begin{align}
 \label{wh-FT}
 &
 \textstyle
 \bar y^r_- = \bar g_r +\bar R_r \bar y^r_+,
\end{align}
As usual \cite{Lee2012PRB,PHe2011JPA},  the term $(1-\bar R_r)^{-1}$ can be decomposed into two parts, $\bar K^r_+(\omega)$ and $\bar K^r_-(\omega)$, namely
\begin{align}
 (1-\bar R_r)^{-1}=\bar K^r_+\bar K^r_-,
\end{align}
where $\bar K^r_\pm(\omega)$ are analytic and nonzero in the upper/lower half-plane with the properties   $\bar K^r_+(\omega)=\bar K^r_-(-\omega)$ and
the term $\bar K^r_\pm(\omega)|_{|\omega|\to\infty}=1$.
Finally,   the eq.(\ref{wh-FT}) can be decomposed  into the following form
\begin{align}
 \label{wh-fc}
 {\bar y^r_{+}}/{\bar K^r_+}=\bar K^r_-\bar g_r - \bar K^r_-\bar y^r_{-}.
\end{align}
Here, term $\bar y^r_+/{\bar K^r_+}$ is analytic in the upper half-plane and term $\bar K^r_-\bar y^r_-$ is analytic in the lower half-plane.

We assume that $\bar K^r_-\bar g_r$ can be further decomposed into two parts
\begin{align}
 \label{wh-dc}
 \bar K^r_-\bar g_r=\bar \varPhi^r_{+}+\bar \varPhi^r_{-},
\end{align}
where $\bar \varPhi^r_+$ and  $\bar\varPhi^r_-$ are analytic in the upper and lower half-plane, respectively. From this equation, we can get
\begin{align}
 &\label{wh-LT}
 {\bar y^r_{+}}/{\bar K^r_{+}}-\bar\varPhi^r_+=\bar\varPhi^r_- - \bar K^r_-\bar y^r_{-}.
\end{align}
In the eq.(\ref{wh-LT}), the left and right hand sides are analytic and bounded in the upper and lower half-planes, respectively.
From Liouville's theorem, both sides of eq.(\ref{wh-LT}) should be a constant.
When $\omega\to\infty$, $\bar y^r_{+}\to0$ and $\bar K^r_\pm\to 1$.
This provides us a way to determine this constant and leads to the solution of $\bar y^r_+$ in the form
\begin{align}
 &
 {\bar y^r_{+}}(\omega)={\bar K^r_{+}}(\omega)[\bar\varPhi^r_{+}(\omega)-\bar\varPhi^r_{+}(\infty)].
\end{align}
With the help of the relations ${\rm e}^{-|\omega|a}=[\alpha/(\epsilon-{\rm i}\omega)]^{{\rm i}\omega a/\pi} [\alpha/(\epsilon+{\rm i}\omega)]^{-{\rm i}\omega a/\pi}$ and $2\cosh \omega a=\sqrt{2\pi} \Gamma(1/2+{\rm i}\omega a/\pi)^{-1} \Gamma(1/2-{\rm i}\omega a/\pi)^{-1}$ where $\epsilon=0^+$ is infinitely small and $\alpha$ is a constant.
Then the factorization of $(1-\bar R_r)^{-1}$ is  given by 
\begin{align}
 \label{wh-kpm}
 \bar K^r_\pm(\pm\omega)=2\pi
 \Gamma\Big(\frac{r}{2w}-\frac{{\rm i}\omega c}{4\pi}\Big)^{-1}
 \Gamma\Big(1-\frac{r}{2w}-\frac{{\rm i}\omega c}{4\pi}\Big)^{-1}
 \Big(\frac{4\pi {\rm e}/c}{\epsilon-{\rm i}\omega c}\Big)^{\frac{{\rm i}\omega c}{2\pi}}.
\end{align}
The functions $\bar K^r_\pm$ is analytic and nonzero in the upper/lower half-plane.
The condition $\bar K^r_\pm(\infty)=1$ can be seen from  the asymptotic form of $\bar K^r_+$ for a large $\omega$:
\begin{align}
 \bar K^r_+(\omega)|_{|\omega c|\gg1}=1
 -\alpha_0 \frac{2\pi}{{\rm i\omega c}}
 +\frac{\alpha_0^2}{2} \Big(2\frac{\pi}{{\rm i\omega c}}
 \Big)^2 +O(\omega^{-3}),
\end{align}
where $\alpha_0=-1/3+r(1-r/2w)/w$.
In the lower/upper half plane, the poles of $\bar K_\pm$ are at $\mp \omega_m$ where
$ \omega_m= 2m\pi {\rm i}/wc, m=1,2,\cdots$.

From the functions $\bar K^r_-$ given in eq. (\ref{wh-kpm}) and  $g_r(k)$  given in eq. (\ref{g}), we may decompose $\bar K^r_-\bar g_r$.
Following \cite{Lee2012PRB}, then the $\bar \varPhi^r_+$ is given by
\begin{align}
 \bar \varPhi^r_+(\omega)
 &
 =\frac{{\rm i}\bar h_r\bar K^r_+(0)}{\omega+{\rm i}\epsilon}
 -\frac{2{\rm i} B_\varepsilon}{wc}\sum_{m=0}^\infty
 \frac{(-1)^m \bar K_+(\omega_m) \bar J_m}{\omega+\omega_m}
 +\bar K_+(\omega) \bar \xi_r-\bar \xi_r,
 \nonumber\\
 \bar J_m
 &
 =-\sum_{r'}U_{rr'} {\rm e}^{{\rm i} \omega_m Q_{r'}} \sin({\rm i} \omega_m r' c/2).
\end{align}
Where $\bar h_r=h\sum_{r'} U_{rr'}$, $\epsilon$ is a infinitely small value and $\epsilon>0$.
By using  the exact form of $\bar \varPhi^r_+$,  the function $\bar y^r_{+}$ is given by
\begin{align}
 \label{wh-ty}
 \bar y^r_{+}&=\bar K^r_+\Big[
 \frac{{\rm i}\bar h_r\bar K^r_+(0)}{\omega+{\rm i}\epsilon}
 -\frac{2{\rm i}B_\varepsilon}{wc}\sum_m\frac{(-1)^m\bar K^r_+(\omega_m)\bar J^r_m}{\omega+\omega_m}\Big]
 +({\bar K^r_{+}}-1)\bar \xi_r.
\end{align}
It is obviously that $\bar y^r_+$ is analytic in the upper half-plane.
The poles in the lower half plane are located at positions $\omega=-{\rm i}\epsilon$ and $\omega_{-m}$, $m=1,2,\cdots$.
With the large $\omega$ asymptotic form of $\bar K^r_+$, the function $\bar y^r_+$ can be expanded to 
\begin{align}
 & \bar y^r_+(\omega)
 = \bar \sigma_1 \frac{2\pi}{{\rm i}\omega c}
 +\bar \sigma_2 \Big( \frac{2\pi}{{\rm i}\omega c}\Big)^2
 +O(\omega^{-3}).
\end{align}
where $|\omega|\gg1$ and
\begin{align}
 &
 \bar \sigma^r_1=-\frac{\bar h_rc}{2\pi}\bar K^r_+(0)
 +\frac{b_\varepsilon}{w\pi}\sum_m(-1)^mK^r_+(\omega_m) \bar J_m-\alpha_0\bar \xi_r,
 \nonumber\\
 &
 \bar \sigma^r_2=\alpha_0\frac{\bar h_rc}{2\pi}
 \bar K^r_+(0)
 -\frac{2b_\varepsilon}{w\pi}\sum_m(-1)^mK^r_+(\omega_m) \bar J_m(\frac{{\rm i}\omega_m c}{2\pi}+\alpha_0)+\alpha_0^2\bar \xi_r/2,\nonumber
\end{align}
The value $y_+(\infty)$ is determined from
$y^r_+(\infty)=-{\rm i}\,\Res[\tilde y^r(0)]=\sum_{r'} U_{rr'}\bar K_{r'}(0)^2\bar h_{r'}=h$, which  is consistent with the boundary conditions from the thermodynamic Bethe ansatz equations.
The other two boundary conditions are
$y^r_+(0)=-{\rm i} \Res[\tilde y^r_+(\infty)]
 =-2\pi \sigma^r_1/c$ and
$y'_+(0)=-\Res[\omega \tilde y^r_+(\omega)|_{\omega=\infty}]
 =(2\pi)^2 \sigma^r_2/c^2$.
We further calculate   $S_\varepsilon$ introduced in (\ref{tba-whp}), i.e.
\begin{align}
 \label{wh-s}
 &
 S_\varepsilon
 =\sum_r\bar J_1^r\Big[\bar y^r_+({\rm i}\omega_0)+\bar \xi_r\Big].
\end{align}
When $hc\ll p$, we have
\begin{eqnarray}
 S_\varepsilon&=&\frac{\pi}{2B_\epsilon(\omega_1^2)}\Big[\frac{w(w^2-1)}{12}h^2+\frac{2}{3}{\pi^2T^2}\Big].
\end{eqnarray}
Then we obtain the low temperature dressed energy $ \varepsilon^{\rm c}$
\begin{align}
 \label{tba-s}
 &
 \varepsilon^{\rm c}
 =k^2-\mu+\alpha b_\varepsilon^{-1} \cosh(\omega_0 k)
 - G_{\rm c}(k,Q_{\rm c})\xi_{\rm c} + \hat G_{\rm c}\bullet\varepsilon^{\rm c}(k)\\
 &
 \alpha=\frac{wc}{2\pi} \Big[\frac{w(w^2-1)}{12}h^2+\frac{w-1}{3}{\pi^2T^2}\Big].
\end{align}
Using these equations, we will calculate the exact result of  low temperature behaviour  for the 1D repulsive Fermi gases with  a weak magnetic  field.

\begin{figure}[ht]
 \includegraphics[width=0.9\linewidth]{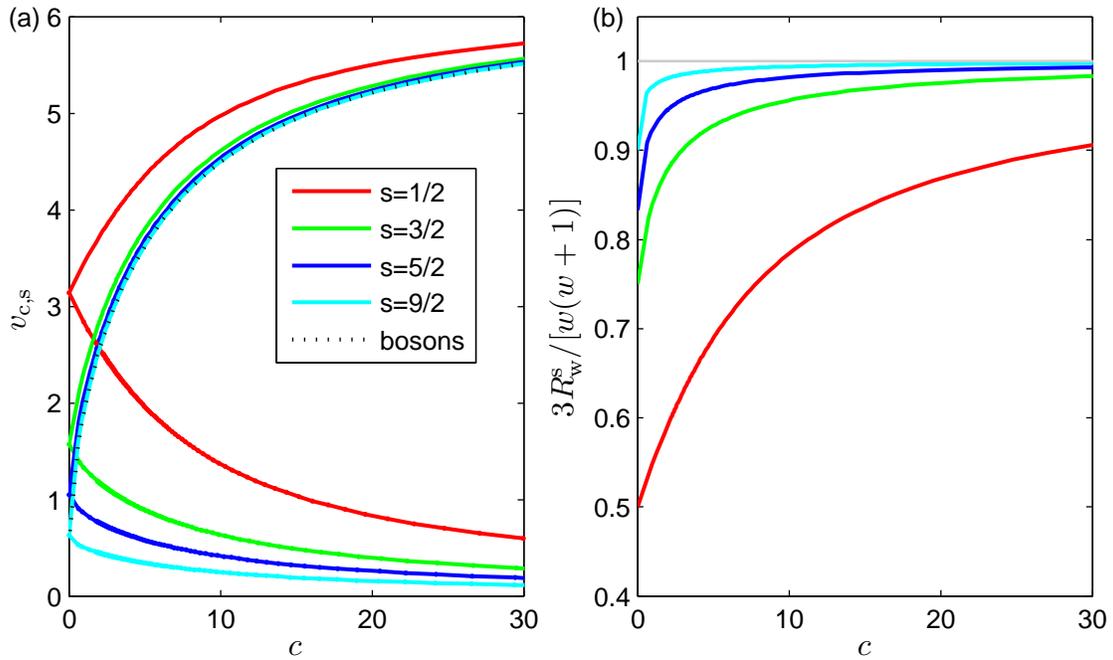}
 \caption{\label{f-vswr}
 Spin and charge  velocities and Wilson ratio for the pseudo spin-$1/2$,  -$3/2$ -$5/2$ and $-9/2$ Fermi gases in weak magnetic field limit, where $ h=0$  and $n=1$. (a) The velocities for  the charge  (upper branch) and the spin (lower branches) degrees of freedom. (b) The Wilson ratio vs interacting strength $c$  for different spin symmetries. It displays  a plateaus of the height $w(w+1)/3$  in the large $w$ limit.  }
\end{figure}

\section{Spin charge separation and low temperature behaviour }
\label{sec-LTSCS}

\subsection{Spin-charge separation for  $SU(w)$ Fermi gases}

At zero temperature, the repulsive $SU(w)$ Fermi gas exhibit an antiferromagnetic ground state.
At low  temperatures,  spin-charge separation behavior naturally  occurs in TLL regime.
The characteristic of the TLL for  this model can be found from the simplified TBA equations (\ref{tba-s}).
The  eq.(\ref{tba-s}) was given  in terms of the chemical potential $\mu$,  the zero dressed energy point $Q_{\rm c}$ at low temperatures   and the  weak magnetic field  $h$.
At $T=0$ and $h=0$, the pseudo-Fermi point is denoted as  $Q_0$.  The integral BA equation in  the charge sector is given by
\begin{align}
\label{bae-r0}
 \rho^{\rm c}_0(k)=\frac{1}{2\pi}+G_{\rm c}|_{Q_0}\bullet \rho_0^{\rm c}(k),
\end{align}
where the corresponding dressed energy is denoted as $\varepsilon^{\rm c}_0(k)$ and the corresponding pressure $p_0(\mu)$ only depends on the chemical potential.
In low temperature limit,  $|Q-Q_0|$ is  very  small and so is their energy difference $\Delta \varepsilon^{\rm c}=\varepsilon^{\rm c}-\varepsilon^{\rm c}_0$, here  $\varepsilon^{\rm c}_0(Q_0)=0$. Up to the leading order, $\Delta \varepsilon^{\rm c}$ satisfies the following equation

\begin{align}
\label{bae-d0}
\Delta\varepsilon^{\rm c}
 =\alpha b_\varepsilon^{-1} \cosh(\omega_0 k)
 - G_{\rm c}(k,Q_{\rm c})\xi_{\rm c}
 + \hat G_{\rm c}|_{Q_0}\bullet \Delta\varepsilon^{\rm c}(k).
 \end{align}
We also can calculate the  pressure via $p=p_0-\hat I\bullet\Delta\varepsilon^{\rm c}/2\pi+\xi_c/\pi$.
Using  eqs. (\ref{bae-r0}) and (\ref{bae-d0}), we have

\begin{eqnarray}
\label{p-h-t}
 p&=&p_0(\mu)-\alpha\frac{b_\rho(\mu)}{b_\varepsilon(\mu)}+2\xi_{\rm c}\rho_0(Q_0)
 =p_0(\mu) +\Delta_h(\mu) h^2 +\Delta_T(\mu) T^2,~~\\
 \label{D-h}
 \Delta_h(\mu) &=&-\frac{w^2(w^2-1)c}{24\pi} \frac{b_\rho }{b_\varepsilon}
 =\frac{w(w^2-1)}{24\pi v_{\rm s}},\\
 \label{D-T}
 \Delta_T(\mu) &=&\frac{\pi^2}{3}
 \Big[\frac{\rho_0(Q_0)}{\varepsilon'(Q_0)}
 -\frac{wc}{2\pi}(w-1) \frac{b_\rho}{b_\varepsilon}
 \Big]
 =
 \frac{\pi}{6} \Big[\frac{1}{v_{\rm c}}+\frac{w-1}{v_{\rm s}}\Big],
\end{eqnarray}
where $b_\rho =\hat I\bullet [\cosh(\omega_0 k) \rho^{\rm c}(k)]$ and
\begin{align}
\label{vs-vc}
 v_{\rm s}
 =\frac{\varepsilon^r(Q_r)'}{2\pi \rho^r(Q_r)}=-\frac{b_\varepsilon}{b_\rho wc},~~~
 v_{\rm c}=\frac{\varepsilon^{\rm c}(Q_{\rm c})'}{2\pi \rho^{\rm c}(Q_{\rm c})}
\end{align}
are the pseudo Fermi velocities in  the spin  and charge  sectors, respectively.
We would like to addressed that the result eq. (\ref{p-h-t})--(\ref{D-T}) are  derived for the first time \footnote{For the $SU(2)$ and $SU(3)$ Fermi gases, similar results were found for strong coupling regions in  \cite{Lee2012PRB,PHe2011JPA}. A discussion on  $SU(w)$ Fermi gases  with $c=\infty$ and $c=0$ was presented in \cite{Schlottmann:1997}.  } and valid for arbitrary interaction strength.
Here the spin velocity $v_{\rm s}=-{b_\varepsilon}/({b_\rho wc})$ can be calculated   from eq.(\ref{tba-s}) with the help of  the Wiener--Hopf solution to the density.
Similarly, using the Wiener--Hopf solution for the charged density, one can also  reach a close form of the sound velocity $v_{\rm c}$.
In the limit $h\to0$, Hamiltonian (\ref{H}) exhibits  the $SU(w)$ symmetry. Therefore all the spin velocities of  $w-1$ branches  are the same under the pure Zeeman splitting.
This means that  the  spin velocity $v_{\rm s}$ in  eq. (\ref{vs-vc}) is independent of  the spin rapidity index `$r$'.
In fact, this property holds for unequal spacing weak Zeeman splittings.

For a weak magnetic field, the density of magnetization and susceptibility are directly reached from eq. (\ref{p-h-t})
\begin{align}
& m_z=\partial_h p=2\Delta_h h,~~
\chi=\partial_h m_z=2\Delta_h=\frac{w(w^2-1)}{12\pi v_{\rm s}}.\label{Un1}
\end{align}
The magnetization linearly responds to the weak external magnetic field with a finite susceptibility.
The susceptibility solely depends on the pseudo spin velocity that presents a universal feature of spin fluctuations.
It is obvious to see  that
\begin{align}
\chi v_{\rm s}={w(w^2-1)}/{12\pi}.\label{Un2}
\end{align}
This provide a universal relation between the  multicomponent spin velocity and magnetic susceptibility. This relation is important to be confirmed in experiment with ultracold fermionic atoms \cite{Pagano2014NP}.
We observed from the FIG. \ref{f-vswr} (a) that spin velocity decreases quickly as we increase the number of components. This means that any  spin flipping process involves all spin states. In this sense, the spin transportation is strongly suppressed for a large number of spin states. 
From FIG. \ref{f-vswr} (a), we also observe that for the large spin system the pseudo charge velocities turns to the pseudo charged velocity of spinless bosonic gas.
This nature agrees with the experimental observation \cite{Pagano2014NP} that the repulsive fermionic gases with $SU(w)$ symmetry exhibit  the bosonic spinless  liquid for  a large value of $w$ at $T=0$ and $h=0$.

From eq. (\ref{p-h-t}), the density of entropy and the specific heat
\begin{align}
\label{se--cv}
& s_{\rm e}=\partial_T p=2\Delta_T T,~~
c_L=T \partial_T s_{\rm e}=2T\Delta_T=\frac{\pi T}{3} \Big[\frac{C_{\rm c}}{v_{\rm c}}+\frac{C_{\rm s}}{v_{\rm s}}\Big],
\end{align}
where $C_{\rm c}=1$ and $C_{\rm s}=w-1$ are the central charges in the charge and spin sectors, respectively.
The low energy properties are uniquely determined by the collective excitations in  spin and charge degrees of freedom. This nature   is called {\it spin-charge} separation that is a  hallmark of   the TLL. 
In the TLL phase, both quantum fluctuation and thermal fluctuation are on equal footing in regard to the temperature.
The dimensionless Wilson ratio between the susceptibility  and the specific heat divided by the
temperature $T$,
\begin{align}
 R^{\rm s}_{\rm w}=\frac43\pi^2\frac{\chi}{c_{\rm L}/T} =
 \frac{\pi v_{\rm c} w(w^2-1)}{3[{C_{\rm c}}{v_{\rm s}}+{C_{\rm s}}{v_{\rm c}}]},\label{Un3}
\end{align}
 measures ratio between the  magnetic fluctuation and the thermal
fluctuation \cite{Wil75,Wan98,XWGuan2013PRL}.
We plot the Wilson ratio against interaction strength for the TLL phase in  FIG. \ref{f-vswr} (b).
This dimensionless ratio displaying plateaus of height $w(w+1)/3$  in the large
$w$ limit, hence  it captures  the spin degeneracy \cite{YCYu2015}.

\subsection{Universality of spin charge separation in quantum gases }

As being discussed  in the  last  section, spin charge separation is one of the most important features of the  TLL.
Since the TLL behavior is the typical nature of 1D quantum gases at  low temperatures, the spin-charge separation may exist  not only in  the weak magnetic field regime but also in a high magnetic field.
This fact is in general true for the integrable quantum gases, for example  $SU(w)$, $Sp(w)$, $SO(w)$  symmetry quantum gases and other  models.
For the 1D integrable quantum Fermi   gases with different high symmetries than  the $SU(w)$, for example,   the $SO(4)$ Fermi gas \cite{Jiang2015}, the low energy spin excitations can be spinon excitations.  In general,  their  TBA equations can be written as
\begin{align}
 \label{g-tba}
 &
 \varepsilon^{{\rm c},\alpha}=\tau_\alpha k^2-\mu_\alpha
 +\sum_\beta \hat K^{{\rm c},\alpha}_{{\rm c},\beta} *\varepsilon^{{\rm c},\beta}_-
 +\sum_n \hat K^{{\rm c},\alpha}_{{\rm s},n} *\varepsilon^{{\rm s},n}_-,
 \nonumber\\
 &
 \varepsilon^{{\rm s},n}= h_n
 +\sum_\alpha \hat K^{{\rm s},n}_{{\rm c},\alpha} *\varepsilon^{{\rm c},\alpha}_-
 +\sum_m \hat K^{{\rm s},n}_{{\rm s},m} *\varepsilon^{{\rm s},m}_-.
\end{align}
Where $\tau_\alpha$ is the single particle dispersion coefficients. The integral kernels satisfy $K^{a,b}_{c,d}=K^{c,d}_{a,b}$.
In order to discuss the low temperature properties, we denote the dressed energies at $T=0$ as $\varepsilon_0$, and $\Delta \varepsilon=\varepsilon-\varepsilon_0$.
With the help of  the Sommerfeld expansion technique, we have
\begin{align}
\label{g-dre0}
 \Delta \varepsilon^{{\rm c},\alpha}=
 &-\sum_\beta K^{{\rm c},\alpha}_{{\rm c},\beta}(k,Q_{{\rm c},\beta})\xi_{{\rm c},\beta}
 -\sum_n     K^{{\rm c},\alpha}_{{\rm s},n}(k,Q_{{\rm s},n})\xi_{{\rm s},n}
 \nonumber\\
 &+\sum_\beta \hat K^{{\rm c},\alpha}_{{\rm c},\beta} \bullet \Delta \varepsilon^{{\rm c},\beta}
 +\sum_n     \hat K^{{\rm c},\alpha}_{{\rm s},n} \bullet \Delta \varepsilon^{{\rm s},n}
 \nonumber\\
 \Delta \varepsilon^{{\rm s},n}
 =&
 -\sum_\alpha K^{{\rm s},n}_{{\rm c},\alpha}(k,Q_{{\rm c},\alpha})\xi_{{\rm c},\alpha}
 -\sum_m      K^{{\rm s},n}_{{\rm s},m}(k,Q_{{\rm s},m})\xi_{{\rm s},m}
 \nonumber\\
 &+\sum_\alpha \hat K^{{\rm s},n}_{{\rm c},\alpha} \bullet\varepsilon^{{\rm c},\alpha}_-
 +\sum_m \hat K^{{\rm s},n}_{{\rm s},m} \bullet\varepsilon^{{\rm s},m}.
\end{align}
At the ground state when magnetic field $h=0$, the integral TBA equations of string densities are
\begin{align}
 \label{g-den0}
 \rho_0^{{\rm c},\alpha}=
 &\frac{\tau_\alpha}{2\pi}
 +\sum_\beta \hat K^{{\rm c},\alpha}_{{\rm c},\beta} \bullet \rho_0^{{\rm c},\beta}
 +\sum_n     \hat K^{{\rm c},\alpha}_{{\rm s},n} \bullet \rho_0^{{\rm s},n}
 \nonumber\\
 \rho_0^{{\rm s},n}
 =
 &\sum_\alpha \hat K^{{\rm s},n}_{{\rm c},\alpha} \bullet\rho_0^{{\rm c},\alpha}
 +\sum_m \hat K^{{\rm s},n}_{{\rm s},m} \bullet\rho_0^{{\rm s},m},
\end{align}
where  $\mu$ is the chemical potential.
From eq. (\ref{g-dre0}) and (\ref{g-den0}), we can find that
\begin{align}
 \sum_\alpha \frac{\tau_\alpha}{2\pi} \hat I\bullet \Delta\varepsilon^{{\rm c},\alpha}
 &=\frac{\tau_\alpha}\pi\sum_\alpha \xi_{{\rm c},\alpha}
 -2\sum_\alpha \xi_{{\rm c},a}\rho^{{\rm c},\alpha}(Q_{{\rm c},\alpha})
 -2\sum_n \xi_{{\rm s},n} \rho^{{\rm s},n}(Q_{{\rm s},n})
 \nonumber\\
 &=\frac{\tau_\alpha}\pi\sum_\alpha \xi_{{\rm c},\alpha}
 -\frac{\pi T^2}{6}\Big[ \sum_a \frac{1}{v_{{\rm c},\alpha}}
 +\sum_n \frac{1}{v_{{\rm s},n}}\Big].
 \nonumber
\end{align}
Up to the order of $T^2$, the pressure is expanded as
\begin{align}
 p &  =-\sum_\alpha \frac{\tau_\alpha}{2\pi} \hat I*\varepsilon^{{\rm c},\alpha}_-=p_0(\mu,h)-\sum_\alpha \frac{\tau_\alpha}{2\pi}\hat I\bullet\Delta\varepsilon^{\rm c,\alpha}+\sum_\alpha \frac{\xi_{\rm c,\alpha}}{\pi}
 \nonumber\\
 &=p_0(\mu,h)+\frac{\pi T^2}{6}\Big[ \sum_a \frac{1}{v_{{\rm c},\alpha}}
 +\sum_n \frac{1}{v_{{\rm s},n}}\Big].\label{p-bose}
\end{align}
The specific heat at  low temperatures is given by
\begin{align}
  c_L=\frac{\pi T}{3}\sum_r \frac{C_r}{v_{r}},~~~C_r=1.
\end{align}
where the sum is carried out  over all the velocities for the charge sector and spin-wave velocities in the spin sector. The central charge for  each branch of the spin degrees of freedom  is $C_r=1$.
Here the  discussion does not depend on the condition that  the  magnetic field is weak or strong.

\section{Equation of state for a strong repulsion}
\label{sec-SR}

In fact the TBA equation of dressed energy  eq. (\ref{TBAr}) or (\ref{tba-s}) can be written as
\begin{align}
\label{tba-st}
 &
 \varepsilon^{\rm c}=k^2-\mu+f(k)+\hat G_{\rm c}*\varepsilon^{\rm c}(k),
\end{align}
where the function $f(k)$ is the contributions from spin  strings.
Here the argument is general true for interacting fermions with high symmetries. When $h\gg T$, the contribution from high spin strings is very small and thus $f(k)=\alpha/b_\varepsilon+O(T^2/c\mu^{3/2}, h^2/c\mu^{3/2}) \approx f_{h\gg T}=-3\alpha \mu^{-3/2}/4$ can be analytically calculated  for TLL regime, also see the attractive case \cite{Guan2010PRA}.
When the external field is very weak, i.e., $h\ll T$, we can find that all the high strings contribute to the pressure, i.e. $f(k)\approx f_{h\ll T}=-T\sinh(wh/2T)/\sinh(h/2T)$ for quantum criticality regime.
At the strong coupling limit, we expand the following  functions which are used in eq. (\ref{tba-s})
\begin{align}
 &
 G_{\rm c}(k)=\sum_{n=0}^{\infty} \frac{g_{2n}}{c^{2n+1}\pi} k^{2n},~~g_{2n}=\frac{(-1)^n}{w^{2n+1}}\Big[\zeta_{2n+1}(1/w)-\zeta_{2n+1}(0)\Big].
\end{align}
Here $\zeta(x)$ is  the Riemann $\zeta$  function. Then the TBA equation is expanded as
\begin{eqnarray}
  \varepsilon^{{\rm c}}(k)&=& \tau k^2-A+O(c^{-5},z^2),~~\tau=1-2g_2p/c^3,
 \\
  A &= &\mu-f_T+\frac{2g_0}{c}p
 -\frac{g_2}{c^3 \pi }\int k'^2\varepsilon^{\rm c }_-(k'){\rm d}k'.
\end{eqnarray}
In the above equations we calculated    $ -\int k^{2n}T\ln\big[1+{\rm e}^{-\frac{\tau k^2-A}{T}}\big]{\rm d}k
={2\Gamma(n+\frac{3}{2})} T^{n+\frac32} {\rm Li}_{n+\frac32}\big[-{\rm e}^{-A/T}\big]/[(2n+1)\tau^{n+1/2}] =-2\Gamma(n+\frac{3}{2}) T^{n+\frac32} F_{n+\frac12}\big(\frac AT\big)/[(2n+1)\tau^{n+\frac12}]$, $\Gamma (5/2)= {3 \sqrt{\pi }}/4,~ \Gamma (3/2)={\sqrt{\pi }}/2$, and  $ \int k'^2\varepsilon^{\rm c}_-(k){\rm d}k =-\frac{\sqrt{\pi }}{2}\frac{T^{5/2}}{\tau^{3/2}}{F}_{3/2}(A/T)$.
In the above equation, $f_T=f_{h\gg T}$ for the TLL phase whereas, $f_T=f_{h\ll T}$ for the quantum critical regime.
Finally,  we obtain the approximation result   of the equation of state
\begin{align}
 & \label{pA}
 p=\frac{\tau^{1/2}}{2\pi^{1/2}} T^{3/2}F_{1/2}\Big(\frac{A}{T}\Big) +O(c^{-5}),\\
 &
 A = \mu-f_T+\frac{2g_0}{c}p
 +\frac{g_2}{c^3 }\frac{T^{5/2}}{2\pi^{1/2}\tau^{3/2} }F_{3/2}(A/T).
\end{align}
This serves as the  equation of state which  describes the exact low temperature thermodynamics of the system for the TLL and quantum critical regime, i.e.
for the regions  $h\ll T$ and   $h  \gg T$, respectively.
%

In the limit $h\gg T$, we find that the Fermi-Dirac integral of eq. (\ref{pA}) describes the low temperature behavior of the TLL.
%
%
In  the strong coupling limit,  the charge and spin  velocities are given by
\begin{align}
 &
 v_{\rm c}=2\pi n(1-4 g_0 n/c),
 ~~
 v_{\rm s}=\frac{4\pi^3n^2}{3wc}(1-6 g_0 n/c),\label{vc-vs}
\end{align}
respectively. Where $g_0=\frac{1}{w}\left(\zeta(\frac{1}{w}-\zeta (0) \right)$. The susceptibility and the specific heat are
\begin{align}
 &
 \chi=\frac{w^2(w^2-1)}{16\pi^4 n^2}(1+6 g_0 n/c),
 \nonumber\\
 &
 \frac{c_{\rm L}}T=\frac{w(w-1)}{4\pi^2 n^2}(1+6 g_0 n/c +2\pi^2n/[3w(w-1)c]).
\end{align}
The Wilson ratio $R^{\rm s}_{\rm w}$ at the strong coupling limit can be expanded as
\begin{align}
 &
 R^{\rm s}_{\rm w}=\frac{w(w+1)}3\Big[1-\frac{2\pi^2 n}{3w(w-1)c} \Big],
\end{align}
see FIG. \ref{f-vswr} (b).  However, for weak coupling limit  $c\to 0$, we can find that
\begin{align}
 v_{\rm s}=\frac{2\pi n}{w},~~~
 \chi=\frac{w^2(w^2-1)}{24 n\pi^2},~~
 \frac{c_{\rm L}}{T}=\frac{w^2}{6n},~~
 R^{\rm s}_{\rm w}=\frac23 w(w-1).
\end{align}


\begin{figure}[ht]
 \includegraphics[width=0.9\linewidth]{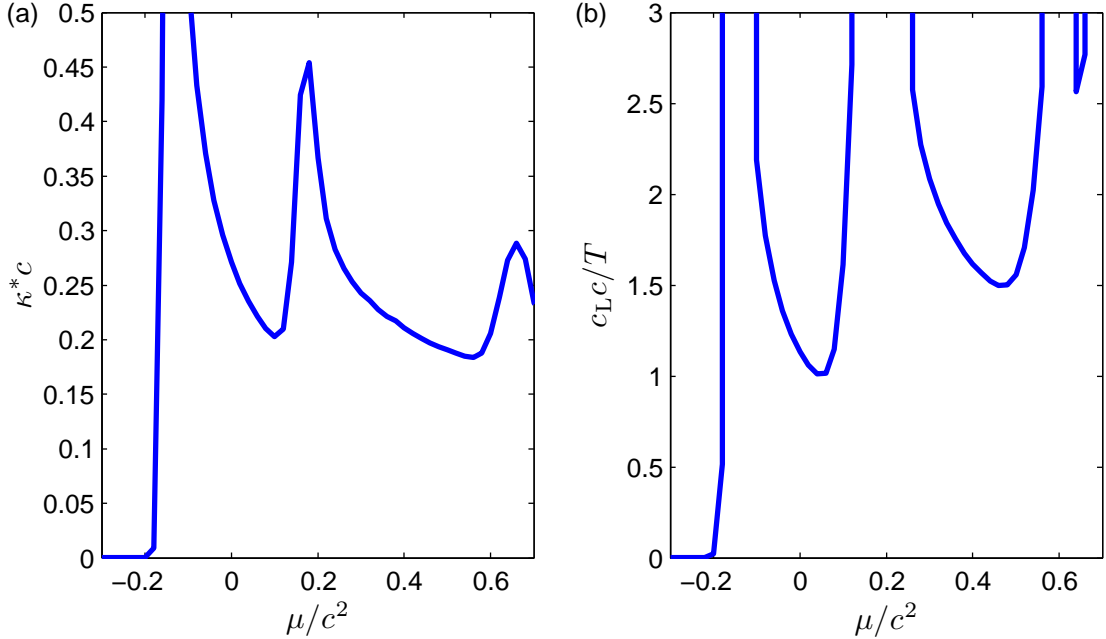}
 \caption{\label{f-ph-t}
 The low temperature behavior of compressibility and specific heat for the spin-$3/2$ Fermi gas at  $T=0.005 c^2$ and $h=0.1 c^2$.
 (a) The dimensionless compressibility. (b) The dimensions specific heat. The round peaks the quantum critical behaviour near the critical point, see a similar  analysis in \cite{GuanXW2011PRA}.}
\end{figure}
The compressibility $\kappa^*$, susceptibility $\chi$ and specific heat $c_{\rm L}$ are divergent at the critical points, see FIG. \ref{f-ph-t} (a) for compressibility and FIG. \ref{f-ph-t} (b) for specific heat, respectively.
These functions have singularities at the critical points and  they  show  universal scaling behavior in the critical regions.
In  the quantum critical regime near the phase transition from vacuum state to the ferromagnetic state,  the  eq. (\ref{tba-st}) and $f(k)=-T\sinh(wh/2T)/\sinh(h/2T)$ give rise to the  quantum criticality.
The equation of state is given by 
\begin{align}
&
p=\frac{T^{3/2}}{2\sqrt{\pi}} F_{1/2} \Big(\frac{A}{T}\Big),~~~
 \frac{A}{T}= \frac{\mu}{T}+\ln\frac{\sinh(wh/2T)}{\sinh(h/2T)}.
\end{align}
It is clear that the the universal scaling function of the pressure is the Fermi-Dirac integral $F_{1/2}$.
In  the quantum critical regime, $h/T\ll1$, thus we have $A/T\approx \ln w+\mu/T+(w^2-1)h^2/(24T^2)$.
Explicitly we have
\begin{eqnarray}
 n&=&\frac{T^{1/2}}{2\pi^{1/2}}F_{-1/2},~~~
 \kappa^*=\frac{T^{-1/2}}{2\pi^{1/2}}F_{-1/2},\\
 m&=&\frac{T^{1/2}}{24\pi^{1/2}}(w^2-1)\frac{h}{T}F_{-1/2},
 \\
  \chi&=&\frac{T^{-1/2}}{24\pi^{1/2}}(w^2-1)
 \Big[F_{-1/2}+\frac{w^2-1}{12}\frac{h^2}{T^2} F_{-3/2}\Big],\\
 s&=&\frac{T^{1/2}}{2\pi^{1/2}}\Big[\frac32 F_{1/2}-\Big(\frac{\mu}{T}+\frac{w^2-1}{12}\frac{h^2}{T^2}\Big) F_{-1/2}\Big],
 \\
 \frac{c_{\rm L}}{T}
 &=&\frac{T^{-1/2}}{2\pi^{1/2}}
 \Big[\frac34F_{1/2}-\frac{\mu}{T}F_{-1/2}
 +\Big(\frac{\mu}{T}+\frac{w^2-1}{12}\frac{h^2}{T^2}\Big)^2 F_{-3/2}\Big],
\end{eqnarray}
where $F_a=F_a(A/T)$ is the Fermi-Dirac integral.
%
%
%

\section{Conclusion}
\label{sec-C}

We have studied  the universal low temperature behaviour of the 1D interacting fermions with $SU(w)$ symmetry via the TBA  method.
We have  developed an analytical method to obtain  the pressure    in terms of charge and spin velocities for the system with arbitrary  interaction strength, see  (\ref{p-h-t})-(\ref{se--cv})  and (\ref{p-bose}).
The low temperature behaviour of these gases which we have  obtained shows a universal   spin-charge separated conformal field theories  of an
effective Tomonaga-Luttinger liquid and an antiferromagnetic $SU(w)$
Heisenberg spin chain.
We have found that the sound velocity  of the  Fermi gases  in the large $w$ limit coincides with that for the spinless Bose gas, whereas the spin velocity vanishes  quickly as $w$ becomes  large, see (\ref{vc-vs}).
In particular,   magnetic properties and the dimensionless Wilson ratio for the high symmetry repulsive Fermi gas  have been derived analytically, see for example, (\ref{Un1})-(\ref{Un3}).
Furthermore, we have studied   the thermodynamics and quantum criticality of the systems  beyond the  regime of the Tomonaga-Luttinger liquid phase in the last section.
These result provides a rigorous   understanding of universal low energy physics of high symmetry interacting fermions  in 1D and sheds light on the experimental  study of  the 1D multicomponent  Fermi gas \cite{Pagano2014NP}.
Moreover, our result will be applicable to the strongly interacting quantum Fermi gases confined in an harmonic potential via the local density approximation.
These strongly interacting systems recently have been received much attention from theory and experiment,  for example, \cite{Volosniev:2014,Levinsten:2015,Dehkharghani,Deuretzbacher:2014,Yangl:2015,Cui:2015,Murmann:2015}.

\section*{acknowledgement}

This work is supported by NSFC (11374331 and 11304357) and by the National Basic Research Program of China
under Grant No. 2012CB922101 and  key NNSFC grant No. 11534014. XWG thank  R J. Baxter and C N Yang for his encouragements and thank  Murray T Batchelor, Angela Foerster,  Jason Ho and Yupeng Wang for helpful discussions.

%

\end{document}